\documentclass[11pt,twoside]{article}
\usepackage{asp2004}
\usepackage{psfig}
\usepackage{epsf}
\usepackage{natbib,graphics}
\usepackage{lscape}
\def\edcomment#1{\iffalse\marginpar{\raggedright\sl#1\/}\else\relax\fi}
\reversemarginpar

\begin{document}
\title{SS\,433: a WR X-ray binary or a WR-type phenomenon ?}
\author{Y. Fuchs, L. Koch Miramond}
\affil{Service d'Astrophysique, CEA/Saclay, Orme des Merisiers b\^at. 709, 91191 Gif-sur-Yvette cedex}
\author{P. \'Abrah\'am}
\affil{Konkoly Observatory, P.O. box 67, 1525 Budapest, Hungary}

\begin{abstract}
   We present mid-infrared spectra 
   of the microquasar SS\,433 obtained
   with the Infrared Space Observatory (spectroscopic mode of ISOPHOT)
   and compare them to the spectra of
   four Wolf-Rayet stars.
   The mid-infrared spectrum of SS\,433 shows
   mainly \ion{H}{i} and \ion{He}{i} emission lines and is very
   similar to the spectrum of WR\,147, a WN8(h)+B0.5V binary with a
   colliding wind. The 2--12\,$\mu$m continuum emission 
   corresponds to optically thin and partially optically thick
   free-free emission from which we calculate a mass loss rate of
   $1.4-2.2\times 10^{-4} M_\odot \,\mathrm{yr}^{-1}$ if the wind is
   homogeneous and a third of these values if it is clumped, which is
   consistent with a strong WN stellar wind. 
	We propose that this strong wind
   outflows from a geometrically thick envelope of material
   surrounding the compact object like a stellar atmosphere, imitating the
   Wolf-Rayet phenomenon. 
\end{abstract}
\thispagestyle{plain}

\vspace*{-0.8cm}
\section{Introduction}
\vspace*{-0.2cm}
	SS\,433 is a microquasar, i.e. an X-ray binary with
	relativistic jets, which shows a very peculiar optical spectrum
	\citep{margon84}:
	there are  the so-called ``stationary'' emission lines, including
	strong H$\alpha$ lines, showing normal Doppler shift movements 
	with a period of 13~days, and 
	the less intense ``moving'' lines
	with huge Doppler shifts
	corresponding to relativistic
	velocities with a period of about 162~days.
	The latter are formed in relativistic jets 
	which undergo a precession
	movement in 162 days. The parameters of this system are:
	$P_\mathrm{orb}=13.08$\,d,
	$P_\mathrm{prec}=162.375\pm0.011$\,d, velocity of the
	ejections $v=0.2647\pm0.0008$\,c, inclination of the jet axis
	to the line of sight $i=78.05^\circ\pm0.5^\circ$ and opening
	angle of the precession cone $\theta=20.93^\circ\pm0.08^\circ$
	\citep{eiken01}.
	The ejections and their precession movement are indeed
	observed in the radio images \citep{hjellming81}. 
	SS\,433 is the only microquasar with continuous ejections and
	which shows evidence of \emph{ions} accelerated to relativistic
	velocities (0.26\,c).
	Despite intensive studies, the nature of the
	two stars in this binary system remains unknown. 
	The presence of Wolf-Rayet-like lines 
	added to the very luminous
	continuum in the visible and near-IR ranges led to associate
	the donor star to a Wolf-Rayet or Of star
	\citep{murdin80,vandenheuvel80}.

\vspace*{-0.3cm}
\section{Observations}
\vspace*{-0.3cm}
	We use observations of SS\,433 	
	(R.A.\,=\,$19^\mathrm{h}11^\mathrm{m}49^\mathrm{s}.57$,   
	Dec.\,=\,$+04^\circ58'57''.8$ in J2000)
	from the archives of the
	Infrared Space Observatory (ISO).
	The observations shown here were achieved
	with the spectroscopic modes of ISOPHOT \citep{lemke96} the photometer on board ISO.
%
	We also searched in the ISO archives for
	observations of WR-type stars, particularly those of WN-type
	which are less evolved and so still emit hydrogen lines. We found
	four WN stars (see
	table~\ref{tabWR}) observed with ISOSWS, 
	and the spectra were degraded to the ISOPHOT spectral resolution
	(FWHM $\sim$\,0.04 \& 0.1\,$\mu$m for the 2--5\,$\mu$m and 
	6--12\,$\mu$m ranges respectively).
	See \citet{fuchs04} for the details about data analysis.
    \begin{table}[!t]
    \caption{Main properties of the WR stars to be compared with
        SS\,433.} 
    \label{tabWR}
   {\small
    \begin{tabular}{@{}l@{\ \ }c@{\ \ }c@{\ }c@{\ }c@{\ \ }l@{}}
\hline
Name & WR type & binary & distance & $A_V$ & reference \\ %
\hline
WR\,78  & WN7h WNL & no & 2.0 kpc & 1.48--1.87 & \citet{crowther95II}\\ 
WR\,134 & WN6 & possible & $\sim$ 2,1\,kpc & 1.22--1.99 & \citet{morel99}
\\
WR\,136 & WN6b(h) WNE-s & possible & 1.8\,kpc & 1.35--2.25 & \citet{stevens99}\\ 
WR\,147 & WN8(h) WNL & B0.5\,V & $630\pm70$\,pc & 11.2 & \citet{morris00}\\ 
\hline
     \end{tabular}}
     \end{table}

\vspace*{-0.cm}
\subsection{The emission lines: comparison with Wolf-Rayet stars}
\begin{figure}[!ht]
	\begin{tabular}{@{}c@{}c@{}}
  \psfig{figure=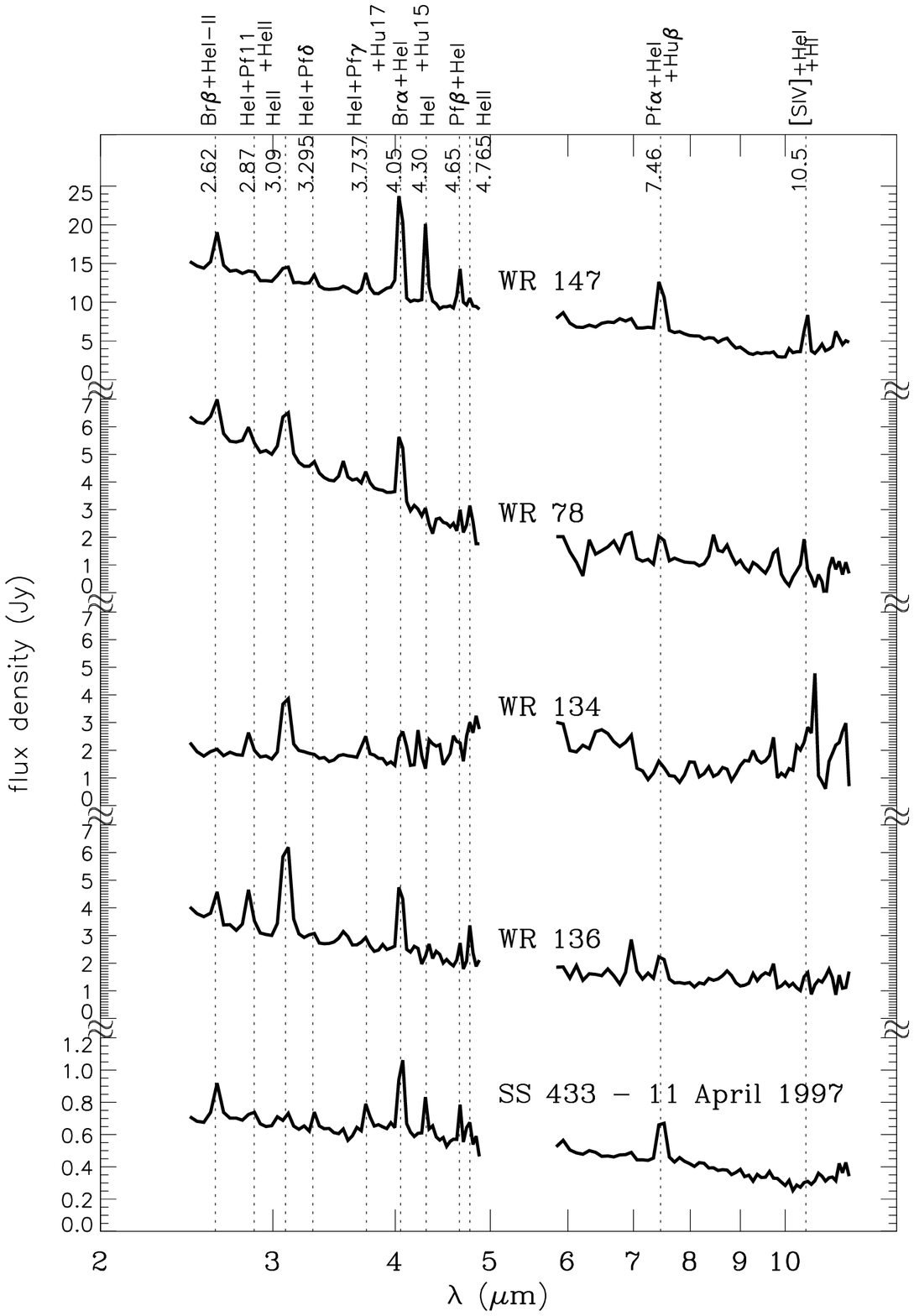,width=6.5cm} &
\psfig{figure=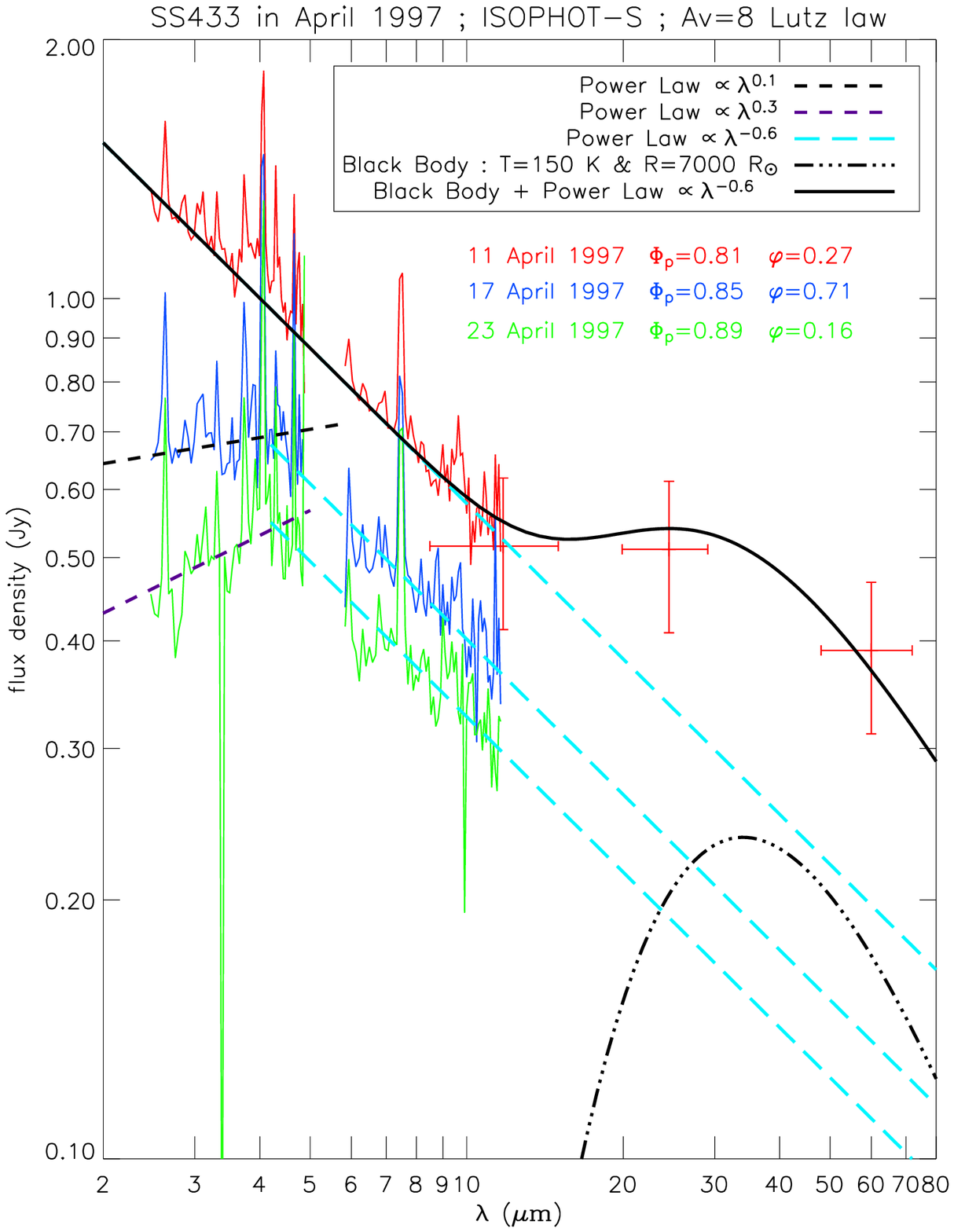,width=6.5cm}
	\end{tabular}
\vspace*{-0.2cm}
   \caption{{\it Left:} Observed (i.e. absorption not corrected) spectra of
   SS\,433 with ISOPHOT-S on 1997 April 11 and of four Wolf-Rayet
   stars with ISOSWS rebined to the PHOT-S resolution and wavelength
   range.
   {\it Right:} Fit of the continuum emission of the dereddened spectra of SS\,433 with power laws and a black body emission (at 4.6\,kpc)  as indicated.}
    \label{figSS433WR}
    \end{figure}

%
%
	Fig.~\ref{figSS433WR} shows the observed spectra (i.e. with no
	correction of absorption by the interstellar medium) of SS\,433
	on 1997 April 11 and of the four WN stars.
%
	No absorption line was found. The emission lines are mainly
	H\textsc{i} lines blended with He\textsc{i} or
	He\textsc{ii} lines. 
	No metallic line is
	detected.  It is clear in Fig.~\ref{figSS433WR} that the
	spectrum of SS\,433 is closest to the one of WR\,147, 
	a WN8(h) star, 
%
	it shows the same
	general shape with the same lines and
	comparable relative intensities between the latters, except
	for the 10.5\,$\mu$m line which is absent in SS\,433. However,
	\citet{smithhouck01} studied 8--13\,$\mu$m spectra of several
	WN8 and WN9 stars where this 10.5\,$\mu$m line can be very weak
	or absent. Thus in the mid-IR SS\,433 looks like a late WN star (WNL)
	of WN8 or later type, which is relatively H-rich.
	Note that WR\,147 is a Wolf-Rayet
	binary (WN8+B0.5V) system with a strong colliding wind that has
	been observed in radio \citep{williams97}.

\subsection{The continuum}
	The spectra were dereddened using the \citet{lutz96} law and $A_V$=8 (see \citealt{fuchs04} for the details).
	The continuum of SS\,433 in April 1997 can be very well fitted
	by one or two power laws and a blackbody emission as shown in
	Fig.~\ref{figSS433WR} (right).  On 1997 April 11 the continuum
	corresponds to a power law from 2.5 to 12\,$\mu$m: 
	$F_\nu = C\,\lambda^{-\alpha}$\,Jy where $C$=2.3 and
	$-\alpha$=$-0.6$. Then a blackbody model with $T$=150\,K
	and $R=7000\,R_\odot$ is added to this power law to
	fit the far-IR emission, but with only three points at 12, 25
	and 60\,$\mu$m the constraints are not strong. The two other April
	1997 spectra are fitted by two power laws
	with spectral index: $-\alpha\!=\!+0.1$ and $-0.6$ on April~17, and 
$-\alpha\!=\!+0.3$ and $-0.6$ on April 23 for the
	\mbox{2--4.5\,$\mu$m} and 
	 4.5--12\,$\mu$m range respectively.
%

	These power laws can be interpreted as free-free emission:
	optically thin for the positive slopes between 2 and
	4.5\,$\mu$m in April 17 and 23, and the negative slope
	($\lambda^{-0.6}$) corresponding to the intermediate regime
	between optically thin and optically thick free-free
	emission. This $-0.6$ slope is the exact theoretical
	spectral index for the free-free emission from an ionized
	homogeneous wind outflowing from a star with a spherical
	expansion and at a constant velocity
	\citep{wrightbarlow75}. This kind of wind is
	very common for O or WR stars \citep{cohen75}. Note that
	\citet{schmid82} demonstrated that the overall spectral
	emitting behaviour is the same for non spherical outflows with
	more complex geometries as long as they stay thick.
%
	The far-IR emission is likely thermal emission from dust
	at $T=150$\,K surrounding the system at a large distance
	\mbox{($R> 7000\,R_\odot$).}

   \subsection{Mass loss evaluation}
	From our fit of the free-free emission of SS\,433 we can
	calculate the corresponding mass loss $\dot{M}$ 
	using the \citet{wrightbarlow75} equation (8). 
	 Except for the flux density at 5\,$\mu$m
	($\nu$=60\,000\,GHz) in the maximum and minimum levels of
	April 1997 ($F_\nu$=876\,mJy \& 495\,mJy) and the distance
	$D$=4.6\,kpc, the parameters of the wind 
	are not very well known. We took
	the usual or average values of the WN winds, assuming that the
	Gaunt factor is $g$=1, the terminal velocity of the wind is
	$\upsilon_\infty$=1000\,km\,s$^{-1}$ \citep{crowther03} and
	using a wind composition typical of late WN stars: mean
	atomic weight per nucleon $\mu$=2, number of free
	electrons per nucleon $\gamma_\mathrm{e}$=1 and mean ionic
	charge $Z$=1 \citep{leitherer97}.
	We find a mass loss rate of
	 $\dot{M}=1.43-2.19\times 10^{-4} \ M_\odot \,\mathrm{yr}^{-1}$.
	We explored the possible range of the parameters to see how
	they influence the resulting mass loss. 
	The greatest uncertainties come
	from the wind velocity and the Gaunt factor so that our result is
	valid within a factor of 2.
	This result is in good agreement with the past
	estimates of $10^{-5}-10^{-4} \ M_\odot \,\mathrm{yr}^{-1}$ \citep{shklovskii81,vandenheuvel81,king00}.

	However, the mass loss rate of SS\,433 has to be corrected
	from the effect due to the very likely inhomogeneity of the
	wind. Clump mass loss rate of WN stars are a factor of 3 times
	lower \citep{crowther03} thus we get
	$4.7-7.3\times 10^{-5} \ M_\odot \,\mathrm{yr}^{-1}$ for SS\,433.
	This result can be compared to the mass loss rate of WR\,147
	measured by \citet{morris00}: $1.5-3.7 \times 10^{-5} M_\odot
	\,\mathrm{yr}^{-1}$ and to the typical range of WN clumped
	mass-loss rates: $10^{-5.5}-10^{-4.5} M_\odot \,\mathrm{yr}^{-1}$ \citep{crowther02}.
	Thus the mass loss rate
	found for SS\,433 is compatible with a strong WN wind.

	With such a huge mass loss 
	one could ask what happens to
	all this ejected material. The relativistic jets of SS\,433
	are well known and the source is continuously ejecting material
	in this way, but \citet{marshall02} and previous studies
	estimated to $\sim10^{-7}\,M_\odot$\,yr$^{-1}$ the
	mass loss of these jets, so 
	negligible compared to the IR wind.
	A part of this wind material probably forms dust at
	$T\sim150$\,K and $R>32.6$\,AU, which then emits in the far
	infrared as we observe for $\lambda > 15\,\mu$m in the 1997
	April 11 spectrum (Fig.~\ref{figSS433WR} right). 
%
	The IR wind is also probably responsible for the
	equatorial outflows observed in radio at larger distances 
	(30--70\,mas $\sim$140--300\,AU, \citealt{paragi99,blundell01})
	perpendicularly to the jet.

\vspace*{-0.3cm}
\section{Discussion}
\vspace*{-0.3cm}
%
%
	We showed that both the emission lines
	and the continuum of the mid-IR spectrum of SS\,433 are
	compatible with a Wolf-Rayet type for the donor star, and
	preferably a late WN star (WN8 or later).
	However, the recent discovery of \citet{gieshuang02} and
	\citet{hillwig04} of absorption features in the blue spectrum of SS\,433 suggests that the donor star is an A3-7\,I
	supergiant star with $10.9 \pm 3.1 \, M_\odot$. 
	On another hand,
	\citet{lopezmarshall04} modelling the
	Chandra X-ray spectrum during eclipse constrain the radius
	of the mass donor star to be $9.1 \pm 1.0 \, R_\odot$ which
	corresponds to O6-O8 main sequence stars with $29 \pm 7 \,
	M_\odot$ and $T=41000-35800$\,K \citep{lang92}. But such a
	radius could also correspond to a B4-5 giant star
	($R=8\,R_\odot$, $M=7\,M_\odot$ and $T=15000$\,K for a
	B5\,III). Note that this constraint on the radius found by
	\citet{lopezmarshall04} is not incompatible with the radius of
	a Wolf-Rayet star, however if the results of \citet{hillwig04}
	are confirmed then the donor star is not a Wolf-Rayet star and
	thus {\bf there is a phenomenon imitating such a star in SS\,433.}

	Before we discuss this point, let us see which constraints on
	the nature of the donor star we can provide from the mid-IR
	spectra of SS\,433. We modeled the possible donor stars with black
	body emissions at a distance of 4.6\,kpc. 
	Both a giant B5 star and a main sequence O6-O8 star have
	negligible emission compared to the optically thin ISOPHOT
	flux density of the continuum and thus they are compatible
	with the spectrum of SS\,433.
	But the emission of an A7I ($T$=8150\,K, $R=60\,R_\odot$) is
	not compatible  unless its
	radius is $<30R_\odot$.
	Then this
	constraint is compatible with the result of \citet{hillwig04}
	who find the radius of the Roche lobe volume for the mass
	donor star to be $R_L = 28 \pm 2 \, R_\odot$. Note that this
	corresponds to a star with an intermediate size between type I
	(supergiant) and type II (bright giant) stars.

%
%
	Now, if the mass donor star of SS\,433 is not a Wolf-Rayet
	star, then there is a phenomenon imitating the emission of
	such a star in this binary system. We propose that this
	phenomenon comes from the material surrounding the compact
	object: this material does not form a classical thin accretion
	disc but rather a thick torus or a thick envelope like a stellar
	atmosphere, which is ionized by the X-rays emitted in the close
	vicinity of the compact object and expelled 
	by radiation pressure, thus imitating the wind of a Wolf-Rayet
	star. 
	There may be a ``classical'' thin accretion disc but only
	very close to the compact object and not detectable because of
	the surrounding material.
	This would explain that there is no evidence for an accretion
	disc in the Chandra X-ray spectrum of SS\,433 according to
	\citet{marshall02} and that SS\,433 is not a strong X-ray
	source as the others microquasars or X-ray transients.
	As Cyg\,X-3 showed very similar results to those of SS\,433
	concerning the IR spectrum and corresponding mass loss
	\citep{koch02} we could ask ourselves about the possibility of
	Wolf-Rayet imitation in this X-ray binary too.

\bibliographystyle{aa}
\bibliography{refss433proc}

\end{document}